\documentstyle[preprint,revtex]{aps}
\begin{document}
\draft
\preprint{To be published in PRB1: Rapid Communications}
\begin{title}
Electromagnetic response of a
static vortex line in a \\ type-II
superconductor :  a microscopic study
\end{title}
\author{Boldizsar Janko and Joel D. Shore}
\begin{instit}
Laboratory of Atomic and Solid State Physics,
Cornell University, Ithaca, New York
\end{instit}
\receipt{  July 1992}
\begin{abstract}
The electromagnetic response
of a pinned Abrikosov fluxoid
is examined in the framework of
the Bogoliubov-de Gennes formalism.
The matrix elements and the
selection rules for both the
single photon (emission -
absorption) and two
photon (Raman scattering)
processes are obtained. The results reveal
striking asymmetries: light absorption by
quasiparticle pair
creation or single
quasiparticle scattering can occur only if the
handedness of the incident radiation is opposite
to that of the
vortex core states.
We show how these effects will lead to
nonreciprocal circular birefringence,
and also predict structure
in the frequency dependence
of conductivity and in the
differential cross section
of the Raman scattering.
\end{abstract}

\narrowtext

Recent far infrared transmission
experiments on superconducting
YBCO thin films \cite{karrai,karrai1}
have revived once again the
interest in the quasiparticle
excitations of the vortex core.
A few decades ago, the microscopic
structure of a vortex line in a
type-II superconductor was the
subject of many theoretical
investigations \cite{3,2,4,9,18,15}.
Despite the rich structure
in the density of states,
predicted by the theory, early
experiments on the microwave
surface resistance \cite{cardona} and
microwave transmission \cite{rose} of
 the mixed state confirmed
only a very simple theoretical
prediction: that the vortex core
states occupy a fraction $ (H/H_{c2})^2 $
of the total volume and
that the vortex line is roughly equivalent
to a normal cylinder of radius
$\sim \xi $,  the superconducting coherence length
. This was largely due to the lack of spatial and
frequency resolution: the experimental signal was
an average over a large number of vortices. More
recently, however,
rather spectacular
results were obtained, by scanning tunneling
microscopy (STM) \cite{hess,hess1} in which
the spatial resolution
of the signal was well
below the coherence length.
These experiments prompted
theoretical work by various groups
\cite{over,shore,gygi} ,
and it turned out, that one can
understand very well the experimental
results within the Bogoliubov
- deGennes (BdG) formalism \cite{2}.

Although the STM experiments reveal
many aspects of the electronic
structure of the vortex, they
do not allow for measurements that
probe the discreteness of the
quasiparticle
spectrum within
the vortex, or the handedness
imposed by the external
magnetic field on the quasiparticle
excitations of the vortex core.
As we will show later on, this
handedness will give rise
to magneto-optical
effects. It was pointed
out to us by Girvin \cite{girvinpc} and
Kapitulnik \cite{kapitpc},
that a high resolution optical
gyroscope \cite{kapit} might be able to
 detect the optical activity
of the vortex core states.
In order to probe the quasiparticle
spectrum, Karra\"i et. al.
\cite{karrai} have recently
performed far
infrared transition
measurements on
$YBa_2Cu_3O_7$
thin films in perpendicular magnetic fields.
They observe a broadened edge-like feature at
 $ 77 cm^{-1}$
and attribute
this to
the
quasiparticle pair creation
process inside the vortex core.
Treating the mixed phase as a
heterostructure, they give a
phenomenological description of their
data, and extract the resonant
frequency mentioned above, which,
however, leads to a somewhat large
 energy gap : $\Delta = 63 meV$.

In this paper we examine the
response of a pinned Abrikosov
vortex to an external electromagnetic
field within the BdG
formalism. We are interested
in the case of an isolated, static
vortex line carrying a single flux
 quantum. The external magnetic
field is taken in the $+z$ direction,
with $H_{c1} < H \ll H_{c2}$.
Due to the cylindrical symmetry,
one can choose the quasiparticle
amplitudes in  the
following form
( {\bf r} = (r, $ \theta $, z )) \cite{3}:
\begin{equation}
\left(
\begin{array}{c}
u_n(r) \\
v_n(r)
\end{array}
\right)
= \frac{1}{\sqrt{2\pi L_z}}
e^{i k_z z}e^{i \mu \theta }
e^{- \frac{i}{2} \sigma _z
\theta }\left(
\begin{array}{c}
g_n^{+}(r) \\
g_n^{-}(r)
\end{array}
\right)\ .\label{sol}
\end{equation}
Here $\sigma _i$ 's
are the usual Pauli matrices. The
magnetic quantum number $\mu $
is half an odd integer.
If we measure the energy
relative to the fermi energy, $E_F$,
and the length in
units of $\xi $, in a gauge where
the pair potential $\Delta (x)$
is real the BdG equations read
\begin{equation}
C \sigma _z \Bigl\lbrace -
\frac{d^2}{dx^2} -
\frac{1}{x}\frac{d}{dx} +
\frac{1}{x^2}\Bigr[ \mu -
\frac{1}{2} \sigma _z \Bigl]^2
 - k_{\rho }^2 \Bigr\rbrace
g_n(r) + \sigma _x
\Delta (x) g_n (x) =
E_n g_n
(x)
\ .
\end{equation}
In the above equation $g$ is a two component spinor,
$C = \hbar ^2 / (2m \xi ^2) $ and $k_{\rho }^2 = k_F^2
- k_z^2$. (For simplicity we have assumed
 an isotropic effective mass.)
In general, a self-consistent
solution to the above equation
can be obtained only numerically
\cite{shore,gygi}. The general
features of the solution are
the following : the bound states, with
exponentially decaying quasiparticle
amplitudes, have an
energy spectrum with
 $E(\mu ) \propto \mu $ at small $\mu $, the
spacing between the levels being $\sim
\Delta_{\infty } ^2 /E_F $ where
$\Delta_{\infty }$ is gap in the
 bulk. The negative
energy states are
fully occupied at $T = 0$, whereas the positive
energy states are empty \cite{9}. The scattering
states have a continuous
spectrum, with energies
$|E| > \Delta _{\infty }$.

The perturbing hamiltonian
describing single-photon
(emission - absorption)
processes, can be given as
\begin{equation}
{\cal H}_{1} = -
\frac{e \hbar}{mci}\int d^3r \sum _{\alpha } \psi
_{\alpha }
^{\dagger } (r) \vec A (r) \vec \nabla \psi _{\alpha } (r)
\ ,
\end{equation}
where the sum is over the spin indices.
The power absorbed by the system
irradiated with an external
electromagnetic
field can be given in two
different ways: ${\cal P} = (2
\omega ^2/c^2) \sigma _1 (\omega )
= \hbar \omega W $ where $\sigma
_1(\omega )$ is the real part of
 the conductivity $\sigma (\omega
)= \sigma _1 (\omega ) + i
\sigma _2 (\omega )$ and W is the
transition rate of the system
under the influence of the
perturbation ${\cal H}_{1}$:
\begin{equation}
W = \frac{2 \pi }{\hbar ^2}
\sum _f |H_{fi}|^2 \delta (E_f - E_i -
\hbar
\omega)
\end{equation}
where $H_{fi} \equiv \langle f|{\cal
H}_1 |i \rangle$.
Depending on the final state $|f \rangle$,
 the matrix elements will
involve different coherence
factors. For example, when the
photon emission - absorption
process scatters a single particle
from one state to another the
transition rate is given as
\begin{equation}
W_{em-abs} = \frac{2 \pi }{\hbar ^2}
\left( \frac{e \hbar }{mc}
\right)^2 \sum _{n,n^{'}}
|M_{n,n^{'}}|^2
\times f_{n^{'}}
\left(1 - f_{n} \right) \delta (E_{n} -
E_{n^{'}} \pm
\hbar
\omega )
\ ,
\end{equation}
where $f_n \equiv f(E_n)$ is
the Fermi function. We consider here
the absorption process only, since
we are ultimately interested
in the low temperature limit,
 where there is no spontaneous photon
emission. The most interesting
case for the present problem is that
of the circular polarized
light : $\vec A(r) = A_q
\hat e_{\pm } \exp(i( \vec q \vec r
- \omega t))$ where $ \vec e_{\pm }$
 stands for the usual
polarization vectors
corresponding
to the left / right
circular polarized light, $\hat e_{\pm } = \hat
e_{x } \pm i\hat e_{y }$ (note that we consider
 the case of a
transverse electromagnetic
wave propagating parallel to, and in the same
direction as, the external magnetic field).
In this particular case the matrix elements are
\begin{eqnarray}
M_{n,n^{'}}^{\pm } = 2 A_q \delta (k_z - k^{'}_z -
q)\delta _{\mu
^{'}, \mu \mp 1} \Big\lbrace
\int dr \left[r g_n^{+} \frac{d}{dr}
g_{n^{'}}^{+} \mp (\mu
^{'} -
\frac{1}{2}) g_n^{+}
g_{n^{'}}^{+} \right]\nonumber \\
- \int dr \left[r g_n^{-}
\frac{d}{dr}g_{n^{'}}^{-} \pm (\mu  +
\frac{1}{2}) g_n^{-}g_{n^{'}}^{-} \right] \Big\rbrace
\ ,
\end{eqnarray}
implying the following selection rules:
\begin{eqnarray}
k_z - k^{'}_z - q = 0 \\
\mu ^{'} - \mu \pm 1 = 0 \\
E_n - E_{n^{'}} - \hbar \omega = 0
\ .
\end{eqnarray}
Further simplification occurs when we
 take the low temperature
limit:
then all the states with
negative angular momentum $\mu^{'} < 0$ are
occupied
\cite{9} and therefore the selection rules
can be satisfied only for
$\mu = \mu ^{'}+ 1$ with
the final state
$\mu = + 1/2 $
and initial state $\mu ^{'} = - 1/2$.
But this is possible {\it only when
the light is left
circularly polarized }.
Explicitly inserting these
values for the angular
momenta, we find
\begin{equation}
M_{n,n^{'}}^{+}
= 2 A_q\int dr r \lbrace g_n^{+}
\frac{d}{dr}g_{n^{'}}^{+} -  g_n^{-}
\frac{d}{dr}g_{n^{'}}^{-} \rbrace
\end{equation}
and
\begin{equation}
M_{n,n^{'}}^{-} \equiv 0
\ .
\end{equation}
Thus, the chirality of the vortex
core states becomes manifest
in the above selection rules
governing the absorption of
circularly polarized
electromagnetic radiation.

The pair creation process,
where the incoming photon creates a pair
of quasiparticle excitations, can be
discussed in a similar
fashion. At zero
temperature only the $\mu > 0$ are available, so
that a photon with an energy just above the pair
 creation threshold
will create a pair
of quasiparticles with opposite spins but the
same angular momentum $\mu = 1/2 $.
The transition probability corresponding to
this process
can be obtained
by explicit calculation in the same way as that of
the single particle process. It is, however,
possible to obtain the
selection rules and the
matrix elements corresponding to this
process directly from the single
particle quantities, via a
particle - hole like
symmetry connecting the positive and negative
energy states \cite{zhangpc}. According to
this symmetry \cite{2}, if
$ (g_n^{+},g_n^{-})$
 describes the state with energy
and momentum $E_n, k_z, \mu$,
then $(g_n^{-}, - g_n^{+})$
corresponds to $ - E_n, - k_z, - \mu$.
In this way,
the single particle
process, where a negative
energy state is destroyed and a positive
energy state is created, can be easily
related to the pair creation
where two positive energy
states are created. In particular, the
selection rule for the angular momentum
for this process is that
$\mu^{'} + \mu = \pm 1$.
Since the pair with the lowest possible
energy has $\mu^{'} = \mu = 1/2 $, the
selection rule can again be
satisfied only by a light
carrying positive helicity, {\it i.e.},
having left circular polarization.

Similar results were obtained
independently, and parallel with our
work, by Zhu, Zhang and Drew \cite{zhou}.
These authors consider
the superconducting
ground state as a vacuum. In this picture,
which is equivalent to ours, the
electromagnetic radiation can be absorbed
only by pair creation at
$T = 0 $. They also point
out that if the carriers in a type II
superconductor are hole type, such as
 in most of the high $T_c$
compounds, the situation
is reversed, with only right
circularly polarized
light is absorbed at zero
temperature.  This is a consequence
of the CT
invariance (simultaneous charge
conjugation and time reversal).

These asymmetries will
probably lead to experimentally
 observable consequences for the
following reason. A difference
in conductivity will cause a
difference in the refractive
index of the system with
respect to the light waves of different
circular
polarization, since the complex refractive
index is given by $N^2
(\omega ) = \epsilon _{\infty }
+ i 4 \pi \sigma (\omega )/\omega
 = \left( n + i\kappa \right) ^2$
where $\epsilon _{\infty }$ is the
 dielectric constant at large
frequencies, $n$ is the real
refractive index and $ \kappa $
is the absorption coefficient.
Different refractive indices
for the two circular polarizations
$n^{\pm }$ will result in a
nonzero Faraday angle, with
 which the polarization
plane of a linearly
polarized light is rotated by a sample with
thickness $z$ : $\phi _F = (\omega z /2c )
(n_{+} - n_{-})$.

An attempt was made by
Karrai et. al. \cite{karrai1} to check the
angular momentum selection rule and look
for chirality in the
vortex  response by
performing circularly polarized light transmission
measurements on superconducting
$YBa_2Cu_3O_7$ thin films.
They found no evidence for
 optical activity in the vortex response.
Instead, the signal is dominated by
magneto-optical effects
attributed to the condensate
and it is interpreted as a cyclotron
resonance of the superconducting
ground state.

A possible explanation for the
lack of optical activity has been
proposed by Hsu \cite{hsu}.
In this alternative picture, the
$ \mu = - 1/2 \rightarrow + 1/2
$ dipole transition is hidden
by the resonance of the circular
motion performed by an {\em unpinned }
vortex, which occurs at
the same
frequency as the dipole
transition $ \sim \Delta _{\infty } ^2 /E_F
$. This resonance is clearly not sensitive
to the polarization of
the external electromagnetic
radiation, which drives the motion of
the vortex. However, we find it
surprising that the resonance of
this circular motion would occur
at such a high frequency.

Let us now discuss the
inelastic light scattering on the vortex
core states.
        The electronic Raman scattering
in metals with energy and
momentum transfer $\omega
= \omega_i - \omega_s $ and $ q = k_i -
k_s$ can be understood as a scattering
on the {\it effective }
density $ \tilde \rho_q
= \sum _{k,\alpha }\gamma_k
a^{\dagger }_{k+q,\alpha }
a_{k,\alpha } $ \cite{genkin,kleindierker}
where the scattering strength $\gamma_k$
is strongly polarization dependent and satisifies
$\gamma_k = \gamma_{-k}$ by time reversal
symmetry. Note that the
continuity equation is
$not$ necessarily valid for $\tilde
\rho_q$ \cite{kosztin}. In direct
space this is equivalent to the
following effective density:
\begin{equation}
\rho _{eff} (r) = \int d^3 R
\sum _ {\alpha }  \psi
_{\alpha }
^{\dagger } (r)
\psi _{\alpha } (r + R ) \gamma (R)
\ .
\end{equation}
For an isotropic system
$\gamma_k = const$ and consequently $
\gamma (R)
= \delta (R) $.  The photon cross
section per unit area and time is
given
by \cite{falkovski}:
\begin{eqnarray}
\FL
\frac{d^2 \sigma }{d \Omega d\omega }
= \frac{1}{2\pi } \left(
\frac{\omega _2}{\omega _1} \right)^2
\frac{1}{|A_{k_i}|^2 |A_{k_s}|^2 \cos \theta }
\sum _f | \langle f|{\cal H}_{eff} |i
\rangle |^2
\delta (E_f - E_i - \hbar \omega )
\ ,
\end{eqnarray}
where $\theta $ is the angle
of incidence and the effective
interaction
hamiltonian is defined as
\begin{equation}
{\cal H}_{eff} =
\frac{r_0}{2}A_{k_i}A_{k_s}
\int d^3 r d^3 r^{'}\sum _ {\alpha }
 e^{i\vec q \vec r}
\gamma ( r^{'} - r )
\psi _{\alpha } ^{\dagger } (r) \psi _{\alpha
} (r^{'} )
\ .
\end{equation}
Here $r_0 = e^2/mc^2 $ is the Thomson radius.
We are now able to examine the matrix
elements corresponding to
different processes. For
example, a photon can be absorbed and
re-emitted by a particle, or a pair
can be created which will
finally recombine into the
condensate and provide the
outgoing photon by emission. The fermion
subspace of the initial and final state
is practically the same as
for the single photon
processes we've discussed.
In the case of an isotropic
system, with $\gamma (r) = \delta (r) $
the
matrix elements can be easily calculated.
Here, for the sake of simplicity,
we choose a special experimental
geometry in which the
direction of the momentum
transfer $q$ is along the $z$ axis. The
generalization for other geometries is
quite straightforward.
The single particle
processes are described by:
\begin{equation}
M_{n,n^{'}} = \int d r r
\lbrace g_n^{+}g_{n^{'}}^{+} -
g_{n}^{-} g_{n^{'}}^{-}\rbrace
\ ,
\end{equation}
along with the selection rules:
\begin{eqnarray}
k_z =  k_z^{'} + q  \\
\mu ^{'} = \mu  \\
E_{n} =  E_{n^{'}} +\hbar \omega
\ .
\end{eqnarray}
The corresponding relations for
the pair creation process
can be obtained with the particle
- hole symmetry mentioned before.

The anisotropic case can also be
discussed if we assume that the
function $\gamma (r) $ obeys
cylindrical symmetry. This is
approximately the case in a
layered superconductor,
when the axis of the
vortices and
the incident and the scattered
light waves are perpendicular to the
layers. Then one can choose the
following trial form : $
\gamma (\vec r) = \gamma _{m,\Gamma }
(r) e^{im\theta }e^{i\Gamma
z}$.
The evaluation of the matrix
elements in this case is somewhat
lengthy
but still straightforward. As a
result we obtain that the selection
rules
are still the same as in the isotropic
case : this is a consequence
of the
cylindrically symmetric
scattering strength $\gamma (r) $.

The more interesting feature is the
dependence of the scattering
cross
section on the transfer energy.
This will be investigated by
evaluating the matrix elements
explicitly, using the numerical
solutions \cite{shore,gygi} of
the Bogoliubov - de Gennes
equations. The detailed
numerical results will be reported
elsewhere. Nevertheless, it is
possible to discuss intuitively
the most important features of
the frequency dependence.
In a bulk superconducting
 phase the dynamic structure factor
is zero until the energy transfer
reaches the $\hbar \omega
= 2 \Delta $ threshold :
the superconducting
ground state cannot
absorb energy unless the photon has enough
energy to break a Cooper pair. Above the
threshold, the dynamic
structure factor is
divergent: the states excluded from the gap
pile up
above $2 \Delta $ giving rise to a square
root singularity in the
density
of states and consequently
in the dynamic structure factor. This
divergency is in practice removed by
gap anisotropy
\cite{kleindierker}
or final state interaction
\cite{monien90}. In a type II
superconductor in the mixed
phase, the situation is quite
different: the
quasiparticle excitations of
the vortex core are capable of
absorbing the
energy of the incoming photon.
  As mentioned before, these states
occupy
a fraction $ (H/H_{c2})^2 $ of the
total volume and provide a
quasi-normal region where
the dissipation is possible {\it unless }
the
energy of the incoming photon is less then
twice the minigap :
$\hbar
\omega < 2 \Delta ^2/ E_F $.
Although this minigap is often very small in
conventional superconductors, it is larger
in a few materials.  For example,
in Nb$_3$Sn, $2 \Delta ^2/ E_F
\sim 0.9 K$  which is is not necessarily
unattainable experimentally
\cite{cardona,greytak}. For high
$T_c$ superconductors, this value
is even higher.

It is also interesting
to ask whether the interaction between the
quasiparticles
would have significant
effect on the Raman spectrum. If the final
state
interaction is included in the description
 of Raman scattering on
 superfluid He II \cite{fredi}
or superconductors \cite{monien90} ,
the
quasiparticles can form a bound state
 below the threshold. Similar effects may
occur in the mixed state of
superconductors.

In conclusion, we have investigated
the electromagnetic response of the superconducting
vortex states. Due to the
handedness of the vortex
states, strong asymmetries are discovered
for the
low temperature absorption of circularly
polarized light :
the $ \mu = -1/2
\rightarrow \mu = 1/2 $ dipole transition and
quasiparticle pair creation are possible {\em only }
with left circular polarization, a striking
consequence of the
handedness
of the vortex core states.
We related the matrix elements to
experimentally accessible quantities
such as the dissipative part
of the
conductivity and
the Faraday angle, and argued that the
asymmetry in
the absorption of circularly polarized
 light in a superconducting
vortex
and a finite rotation of the
plane of polarization of a linearly
polarized light at low temperatures
might be experimentally
observable.

We proposed that inelastic
light (Raman) scattering could be used to
investigate the vortex core states and
developed the
theory of Raman
scattering on a superconducting vortex
for isotropic and two dimensional, layered
superconductors.
We argued that the presence of the vortex
core states will lead to
finite
Raman intensity below
the usual $2 \Delta$ threshold and
that the minigap in the discrete
spectrum of the vortex core
states may be observable using low frequency
Raman spectroscopy. This is even more
likely in high $T_c$
superconductors, where
such a measurement, if successful, can
reveal the magnitude of the energy gap.
The results of conventional
measurements for the energy
 gap in the cuprates, such as tunneling,
infrared and high frequency Raman
spectroscopy are not yet
consistent,
mainly due to the fact
that at high frequencies the spectrum is
convolved
with the phonon and other excitation
spectra \cite{sievers}.

We are grateful to  S.M. Girvin,
A. J. Sievers and especially to
V. Ambegaokar for many useful
discussions and comments.
We are indebted to F.-C. Zhang for
sending a preprint of Ref.
\cite{zhou} prior to publication and for
a very meaningful exchange of ideas.
We would also like to thank J. P. Sethna
for the academic and financial support
through the NSF Grant
No. DMR-9118065.

\end{document}